%% file: main.tex
\documentclass[pdflatex,sn-mathphys-num]{sn-jnl}

\usepackage{graphicx}%
\usepackage{multirow}%
\usepackage{amsmath,amssymb,amsfonts}%
\usepackage{amsthm}%
\usepackage{mathrsfs}%
\usepackage[title]{appendix}%
\usepackage{xcolor}%
\usepackage{textcomp}%
\usepackage{manyfoot}%
\usepackage{booktabs}%
\usepackage{algorithm}%
\usepackage{algorithmicx}%
\usepackage{algpseudocode}%
\usepackage{listings}%
\usepackage{subcaption}
\usepackage{rotating}

\theoremstyle{thmstyleone}%
\theoremstyle{thmstyletwo}%

\theoremstyle{thmstylethree}%

\newif\ifshowred
\showredtrue   

\begin{document}

\title[Article Title]{Scalable AI-based clinical communication training and automated assessment}

\author[1]{\fnm{Masum} \sur{Hasan}}\email{m.hasan@rochester.edu}

\author[2]{\fnm{Ron} \sur{Epstein}}\email{ronald\_epstein@urmc.rochester.edu}

\author[2]{\fnm{Thomas} \sur{Carroll}}\email{thomas\_carroll@urmc.rochester.edu}


\author*[1]{\fnm{Ehsan} \sur{Hoque}}\email{mehoque@cs.rochester.edu}

\affil[1]{\orgdiv{Department of Computer Science}, \orgname{University of Rochester}, \orgaddress{\street{250 Hutchison Rd}, \city{Rochester}, \postcode{14620}, \state{NY}, \country{USA}}}

\affil[2]{\orgdiv{University of Rochester Medical Center}, \orgaddress{\street{601 Elmwood Avenue}, \city{Rochester}, \postcode{14642}, \state{NY}, \country{USA}}}

\abstract{
\input{sections/0._abstract.tex}
}




\maketitle

\input{sections/1._introduction.tex}

\input{sections/2._results.tex}

\input{sections/3._discussions.tex}

\input{sections/4._methods.tex}

\backmatter




\begin{appendices}
\input{sections/z._appendix.tex}

\end{appendices}


\bibliography{ref}

\end{document}

%% file: sections/0._abstract.tex

Poor clinical communication can delay care, contribute to errors, and harm patients, yet opportunities for repeated practice with feedback remain limited. Our prior randomized trial showed that practice with the SOPHIE AI patient platform improved serious illness communication, but the system addressed a single clinical context and required human effort for delivery and assessment. We developed SOPHIE 2.0, a browser-based, self-service platform integrating embodied AI-patient interactions, personalized feedback, and automated assessment across 24 clinical scenarios. An automated large language model assessor evaluated three communication skills---Empower, Be Explicit, and Empathize---with agreement comparable to individual human raters ($r=0.759$; ICC$=0.746$). In a study of 59 clinicians and students, participants completed two AI-patient encounters with personalized feedback; 92\% found the platform engaging, 86\% easy to use, and 83\% clinically relevant. Scores were higher in the second encounter, though the uncontrolled design precludes attributing this change specifically to training.


%% file: sections/1._introduction.tex
\section{Introduction}
\label{sec:introduction}

\begin{figure}
    \centering
    \includegraphics[width=\linewidth]{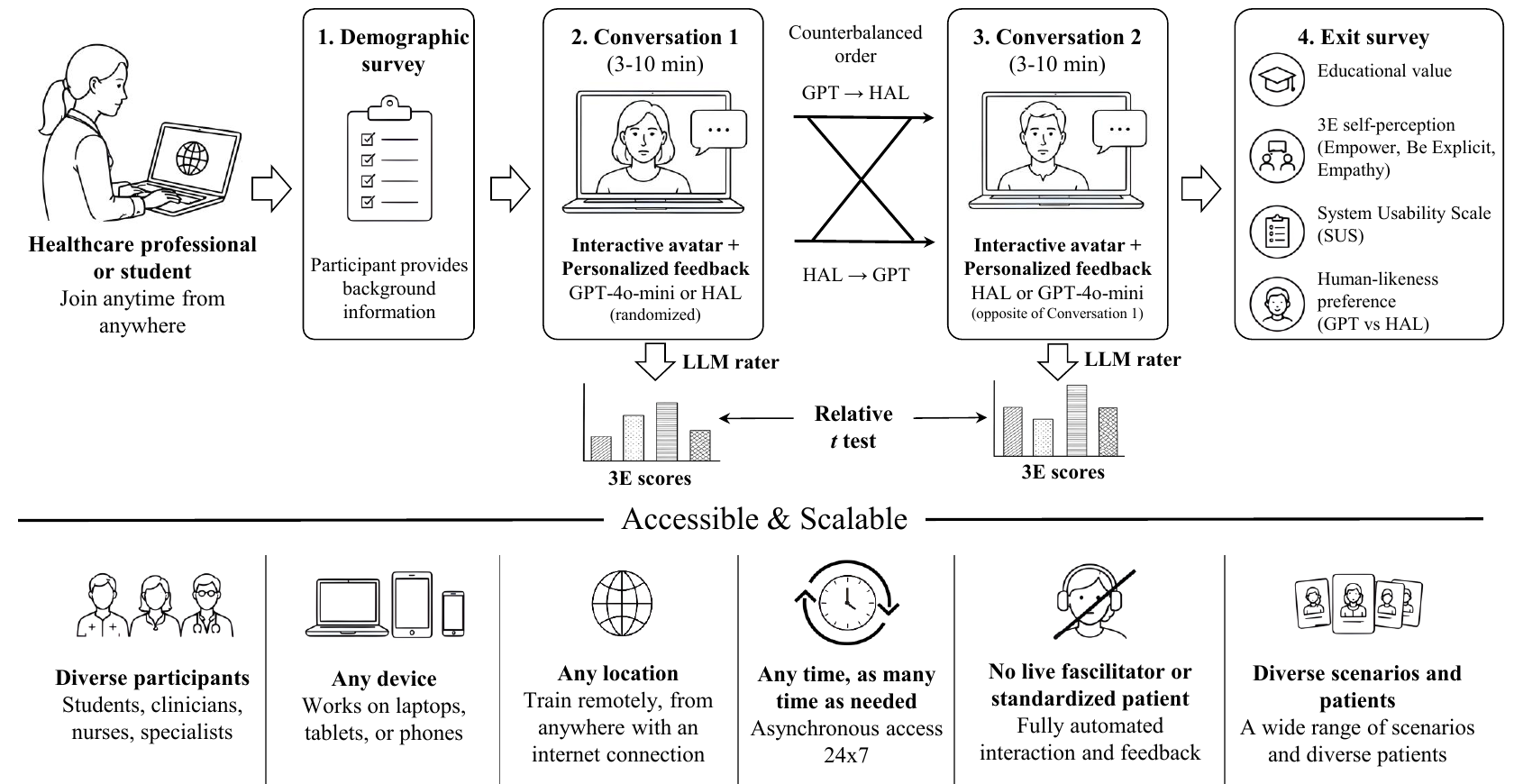}
    \caption{\textbf{Study design of SOPHIE 2.0}, Fifty-nine participants completed a demographic survey, two counterbalanced AI-patient conversations powered by HAL~\cite{hal} and GPT-4o-mini~\cite{gpt} with personalized feedback, and an exit survey. Conversations were automatically evaluated by an LLM rater for 3E communication skills \cite{mvp} and analyzed. The study examines short-term performance, usability, and perceived educational value of a scalable, accessible, and fully automated browser-based platform.}
    \label{fig:teaser}
\end{figure}

Clinical communication is not ancillary to care; it shapes care. In clinical communication, clinicians must explain uncertainty, elicit what matters to patients, and translate those values into a plan. Poor communication among clinicians and between clinicians and patients is associated with delayed care, diagnostic and treatment errors, physical harm, prolonged hospitalization, and lower patient satisfaction \cite{poor1,poor2,poor3}. Developing these skills requires deliberate, repeated practice with timely, specific feedback \cite{feedback1,feedback2,feedback3}. For many learners, such practice remains scarce.

Standardized patients (SPs) provide realistic encounters and actionable feedback and remain a cornerstone of communication training \cite{sp1,sp2,sp3}. However, SP programs are often limited by cost, scheduling, location, limited demographic diversity, and variation among actors, which together constrain both access and repetition \cite{sp-limit1,sp-limit2,sp-limit3,sp-limit4}.

AI-driven virtual patients offer one route towards this goal. Past work has shown that clinical communication behaviors can be quantified using structured rubrics \cite{ali2021novel}, and avatar-based simulations can be used in specific clinical situations \cite{haut2023validating,poker,dental,vsp1,vsp2}. Our previous randomized controlled trial (RCT) evaluated SOPHIE (referred here as SOPHIE 1.0 for clarity) \cite{sophie1.0}, a platform where participants spoke with a generative AI patient and received clinically grounded feedback on three communication skills: Empower, Be Explicit, and Empathize, collectively termed the 3Es \cite{mvp}. However, this study was limited to a single clinical scenario (i.e., an older female patient with advanced cancer considering treatment options), relied on human operators and expert standardized patient raters for evaluation. It therefore established efficacy without resolving whether training, feedback, and assessment could be delivered unsupervised and independently across a broader range of clinical situations.

SOPHIE 2.0 was developed to address this implementation gap. The platform provides a fully automated, browser-based workflow through which learners can select a scenario, complete a face-to-face encounter with an embodied AI patient, receive immediate personalized feedback, and repeat practice without a live facilitator, SP, or human rater (Fig.~\ref{fig:teaser}). Its case library expands the original single scenario to 24 clinical scenarios spanning cardiology, primary care, oncology, critical care, pediatrics, neonatology, surgery, dental, and oral health care. The cases involve virtual patients of a wide range of ages, genders, ethnicities, and appearances. We also developed an automated large language model (LLM) assessor for the 3E framework to reduce the expert effort required to evaluate each encounter. Together, these features address major barriers to broader implementation, including personnel requirements, assessment effort, scheduling, location, and scenario availability.

The present study examined whether the efficacy demonstrated with SOPHIE 1.0 could be translated into a self-service model that reduces the human-resource requirements of communication training. First, we evaluated whether an automated LLM assessor could reproduce human ratings of communication behavior. Second, we evaluated whether healthcare professionals and students could independently complete successive SOPHIE 2.0 encounters and receive automated personalized feedback across a broader range of clinical scenarios. As a secondary analysis, we examined whether the previously observed text-based human-likeness advantage of our custom language model, HAL~\cite{hal}, persisted in an embodied audiovisual interaction. Together, these analyses test whether clinical communication simulation, assessment, and personalized feedback can be integrated into an on-demand platform without requiring standardized patients, facilitators, or human raters.

%% file: sections/2._results.tex
\section{Results}
\label{sec:results}

\subsection{Study participants}

\begin{figure}
    \centering
    \includegraphics[width=\linewidth]{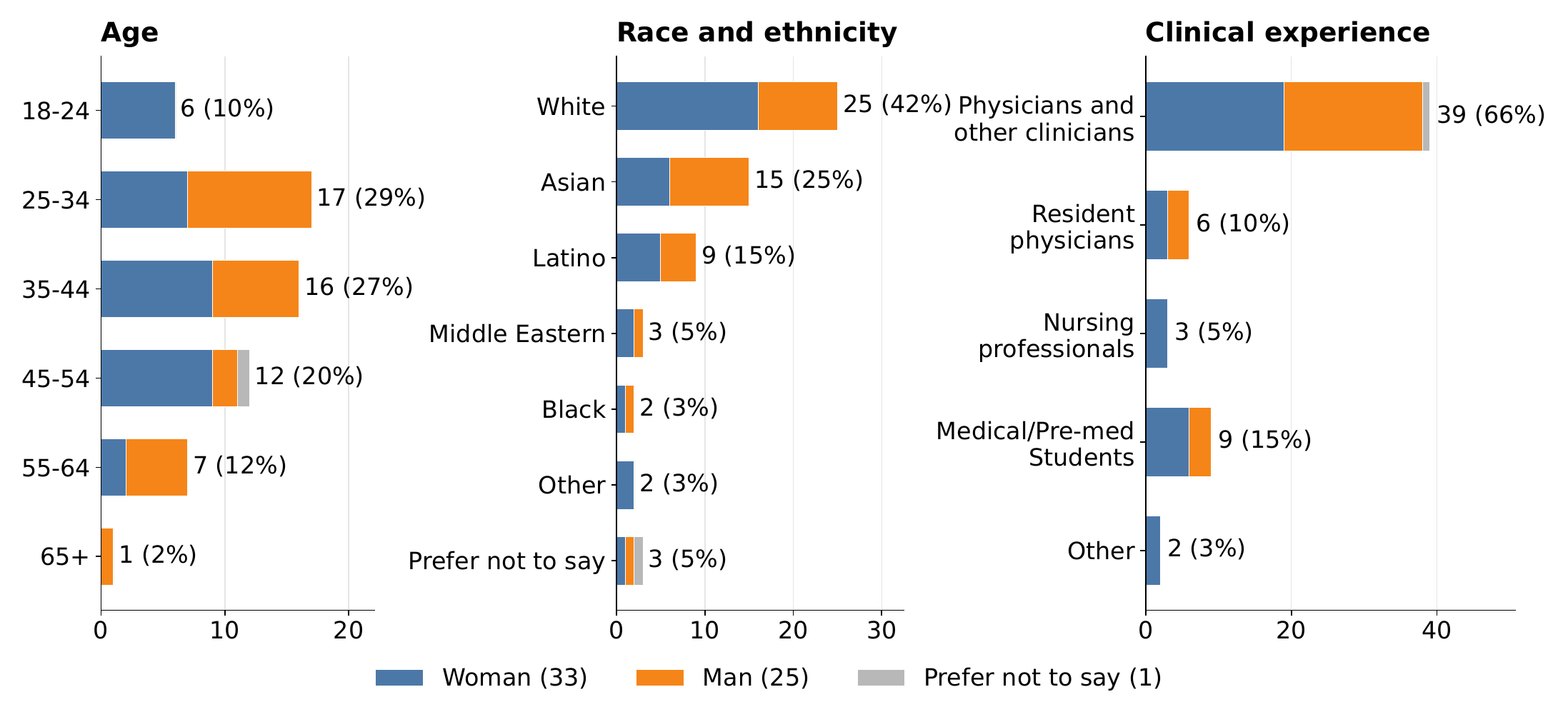}
    \caption{Demographic distribution of $N=59$ participants.}
    \label{fig:demographic}
\end{figure}

The study was conducted between February 2026 and July 2026. During this time period, 91 individuals completed the demographic survey. Among them, 65 (71\%) initiated at least one SOPHIE interaction and 59 (65\%) completed both AI-patient conversations and the exit survey. These 59 participants comprised the primary analytic sample; 34 (58\%) interacted with the GPT-4o-mini, then HAL (GPT-4o-mini$\rightarrow$HAL), and 25 (42\%) completed HAL and then GPT-4o-mini (HAL$\rightarrow$GPT-4o-mini). 8 participants raised technical issues using our built-in reporting system, which were resolved by the authors immediately, and 7 of those participants completed the study.

The participants ranged in age from 18--24 years to 65 years or older, with 45 (76\%) between 25 and 54 years of age. The sample included 33 women (56\%), 25 men (42\%), and one participant (2\%) who preferred not to state their gender. Participants identified as White (25, 42\%), Asian (15, 25\%), Latino (9, 15\%), Middle Eastern (3, 5\%), Black (2, 3\%), or another race or ethnicity (2, 3\%); three participants (5\%) preferred not to state their race or ethnicity. The cohort included 39 physicians and other clinicians in Family Medicine, Palliative care, Gynecology, Dentistry, and others (66\%), six resident physicians (10\%), three nursing professionals (5\%), nine medical or pre-medical students (15\%), and two participants in other roles (3\%) (Fig.~\ref{fig:demographic}).

\subsection{LLM-based assessment of clinical communication aligns with human raters}
\label{sec:llm-judge}

Agreement with the mean human ratings varied across models and communication dimensions (Table~\ref{tab:llm_validation}a). Correlations were generally strongest for Empower and Empathy and weaker for Explicit, suggesting that explicit communication behaviors were more difficult to evaluate consistently. Across the candidate models, the highest Pearson correlations were 0.799 for Empower, 0.598 for Explicit, 0.698 for Empathy, and 0.774 for the overall 3E score. Based on its combined performance across dimensions and its low prediction error, GPT-5.4 with low reasoning was selected as the automated judge for the subsequent analyses. This model achieved an overall Pearson correlation of 0.759 and the lowest overall mean absolute error (MAE) of 0.106.

We next evaluated how the selected LLM judge compared with individual human raters using a leave-one-rater-out analysis (Table~\ref{tab:llm_validation}b). Relative to the consensus of all human rating sources, the LLM achieved a Pearson correlation of 0.759, a Spearman correlation of 0.720, an MAE of 0.106, and an ICC(A,1) of 0.746 (95\% CI, 0.640--0.830). Its performance remained similar when the SP ratings were excluded from the reference consensus, with a Pearson correlation of 0.719 and an ICC(A,1) of 0.702 (95\% CI, 0.570--0.800).

Across the six rating sources, the LLM ranked third for both Pearson and Spearman correlations, achieved the lowest MAE (0.106), and had the second-highest ICC (0.746), closely following TP0 (0.780) and TP2 (0.744). Its performance, therefore, fell within the upper range of the human raters rather than outside their natural variability.


\begin{table}[t]
\centering
\caption{
\textbf{Validation of the LLM judge for communication skill rating.}
In \textbf{a)}, we benchmark different LLM judges against human mean ratings. Higher is better for Pearson; lower is better for MAE.
In \textbf{b)}, we compare the best LLM judge from \textbf{a)} (GPT-5.4 low reasoning), and all human judges in a leave-one-rater-out manner.
}
\label{tab:llm_validation}

{\fontsize{8}{8}\selectfont

\setlength{\tabcolsep}{3pt}
\begin{tabular}{l c cc cc cc cc}
\multicolumn{10}{l}{\textbf{a)} \quad Benchmarking LLM judges} \\
\toprule
\addlinespace[2pt]

\textbf{Model} & \textbf{Reasoning} 
& \multicolumn{2}{c}{\textbf{Empower}} 
& \multicolumn{2}{c}{\textbf{Explicit}} 
& \multicolumn{2}{c}{\textbf{Empathy}} 
& \multicolumn{2}{c}{\textbf{All}} \\
\cmidrule(lr){3-4}
\cmidrule(lr){5-6}
\cmidrule(lr){7-8}
\cmidrule(lr){9-10}

&
& \textbf{Pearson} & \textbf{MAE} 
& \textbf{Pearson} & \textbf{MAE} 
& \textbf{Pearson} & \textbf{MAE} 
& \textbf{Pearson} & \textbf{MAE} \\
\midrule

GPT-4.1 & - & 0.777 & 0.119 & \textbf{0.598} & \textbf{0.138} & 0.697 & 0.174 & \textbf{0.774} & 0.108 \\
GPT-4.1-mini & - & 0.666 & 0.131 & 0.495 & 0.146 & 0.626 & 0.168 & 0.710 & 0.140 \\
GPT-5 & low & 0.746 & 0.111 & 0.484 & 0.173 & \textbf{0.698} & 0.138 & 0.773 & 0.129 \\
GPT-5 & high & 0.732 & 0.122 & 0.466 & 0.178 & 0.670 & 0.142 & 0.752 & 0.133 \\
GPT-5-mini & low & 0.739 & 0.116 & 0.484 & 0.149 & 0.578 & 0.187 & 0.714 & 0.115 \\
GPT-5-mini & high & 0.761 & 0.131 & 0.477 & 0.216 & 0.653 & 0.156 & 0.761 & 0.154 \\
GPT-5.4 & low & \textbf{0.799} & \textbf{0.098} & 0.426 & 0.170 & 0.670 & \textbf{0.136} & 0.759 & \textbf{0.106} \\
GPT-5.4 & high & 0.783 & 0.128 & 0.414 & 0.160 & 0.661 & 0.137 & 0.746 & 0.126 \\
GPT-5.4-mini & low & 0.665 & 0.130 & 0.471 & 0.152 & 0.509 & 0.230 & 0.641 & 0.140 \\
GPT-5.4-mini & high & 0.759 & 0.127 & 0.452 & 0.149 & 0.636 & 0.151 & 0.732 & 0.116 \\

\bottomrule
\end{tabular}

\vspace{8pt}

\setlength{\tabcolsep}{4pt}
\begin{tabular}{llccclr}
\multicolumn{7}{l}{\textbf{b)} \quad Comparing judges in leave-one-out manner} \\
\toprule
\addlinespace[2pt]

\textbf{Rater} &
\textbf{Reference consensus} &
\textbf{Pearson$\uparrow$} &
\textbf{Spearman$\uparrow$} &
\textbf{MAE$\downarrow$} &
\textbf{ICC(A,1)} &
\textbf{95\% CI} \\
\midrule

LLM & SP, TP0, TP1, TP2, TP3 
& 0.759 & 0.720 & 0.106 & 0.746 & [0.640, 0.830] \\


SP & TP0, TP1, TP2, TP3 
& 0.680 & 0.667 & 0.155 & 0.633 & [0.490, 0.740] \\

TP0 & SP, TP1, TP2, TP3 
& 0.806 & 0.802 & 0.111 & 0.780 & [0.680, 0.850] \\

TP1 & SP, TP0, TP2, TP3 
& 0.625 & 0.559 & 0.145 & 0.596 & [0.430, 0.720] \\

TP2 & SP, TP0, TP1, TP3 
& 0.825 & 0.810 & 0.128 & 0.744 & [0.620, 0.830] \\

TP3 & SP, TP0, TP1, TP2 
& 0.642 & 0.582 & 0.126 & 0.638 & [0.500, 0.740] \\

\bottomrule
\end{tabular}

}
\end{table}



\subsection{Participants perceived SOPHIE as educationally valuable and usable}
\label{sec:educational-sus}

\begin{figure}
    \centering

    \begin{subfigure}[t]{0.85\textwidth}
        \centering
        \includegraphics[width=\linewidth]{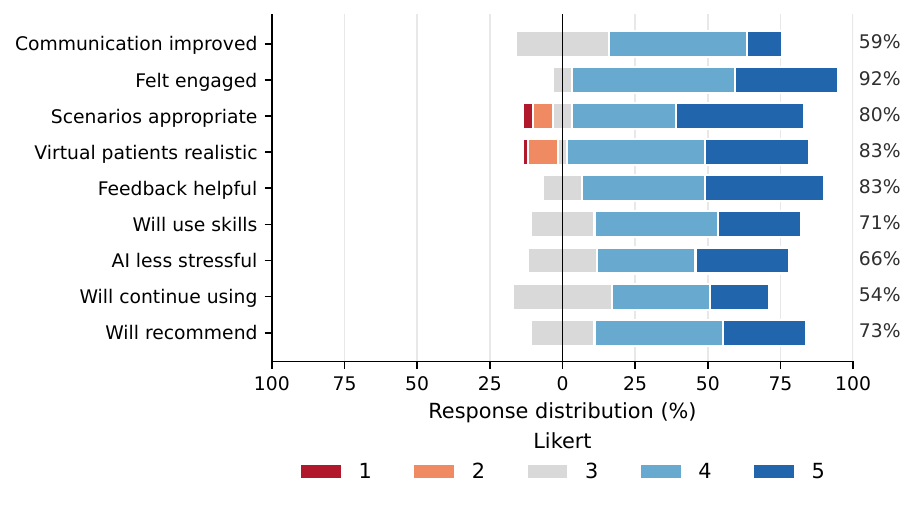}
        \caption{SOPHIE's educational value related questions. Here, the scale ranges from 1 = ``Strongly disagree" to 5 = ``Strongly agree".}
        \label{fig:educational_value}
    \end{subfigure}

    \vspace{1em}

    \begin{subfigure}[t]{0.85\textwidth}
        \centering
        \includegraphics[width=\linewidth]{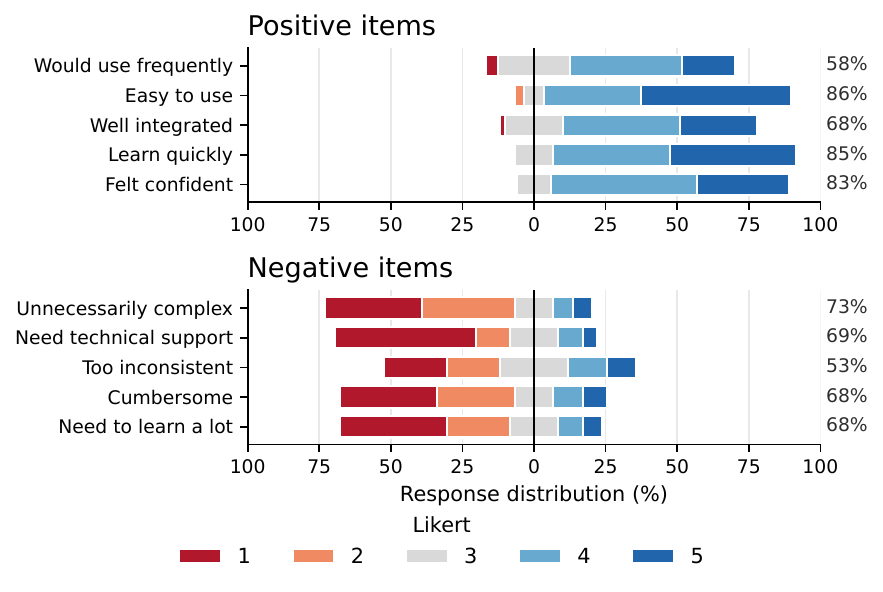}
        \caption{Participant responses to the System Usability Scale (SUS). 1 = ``Strongly disagree" to 5 = ``Strongly agree".}
        \label{fig:sus}
    \end{subfigure}

    \vspace{1em}
    
    \begin{subfigure}[t]{0.45\textwidth}
        \centering
        \includegraphics[width=\linewidth]{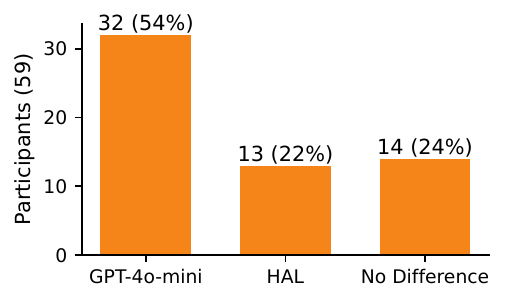}
        \caption{Natural/human-like LLM.}
        \label{fig:more_natural_llm}
    \end{subfigure}
    \hfill
    \begin{subfigure}[t]{0.45\textwidth}
        \centering
        \includegraphics[width=\linewidth]{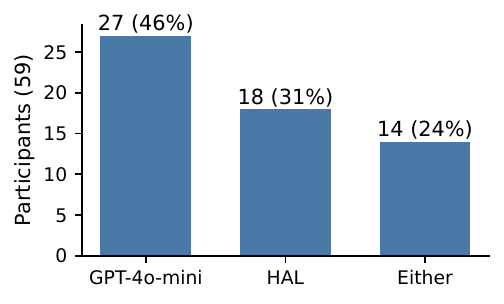}
        \caption{Preferred LLM.}
        \label{fig:preferred_llm}
    \end{subfigure}
    
    \caption{
    Post-completion survey results assessing participants' perceptions and
    preferences regarding the two LLMs, the perceived educational value of
    SOPHIE, and system usability. Sub-figures \textbf{(c)} and \textbf{(d)}
    show divergent Likert bar charts. The percentage indicates the proportion
    of responses representing the desired outcome: 4 or 5 for positive items,
    and 1 or 2 for negative items.
    }
    \label{fig:survey_results}
\end{figure}

The educational-value survey assessed whether SOPHIE was engaging, clinically relevant, and helpful for practicing communication skills. As shown in Figure~\ref{fig:educational_value}, 92\% of participants felt engaged, 80\% considered the scenarios appropriate for their clinical roles, and 83\% reported that the virtual patients elicited skills used in real clinical communication. Personalized feedback was considered helpful by 83\%, while 71\% anticipated applying the skills they practiced, and 73\% would recommend SOPHIE. 
Full questionnaire provided in supplementary materials.

Figure~\ref{fig:educational_value} also shows that 59\% of the participants perceived an improvement in their communication skills;  Figure~\ref{fig:3e_self_perception_distribution} provides a further detailed breakdown of that. 66\% found interacting with AI less stressful than interacting with a person, and 54\% intended to continue using SOPHIE. 

The System Usability Scale items assessed whether participants could use SOPHIE frequently, confidently, and independently. As shown in Figure~\ref{fig:sus}, 86\% found the platform easy to use, 85\% believed it could be learned quickly, 83\% felt confident using it, and 58\% reported that they would use it frequently. Most participants also rejected the need for technical support and descriptions of the platform as complex or cumbersome. Overall, SOPHIE appeared easy to learn and use with limited support. 

\subsection{Participants reported favorable communication self-perceptions}

\begin{figure}
    \centering

    \begin{subfigure}[t]{0.70\textwidth}
        \centering
        \includegraphics[width=\linewidth]{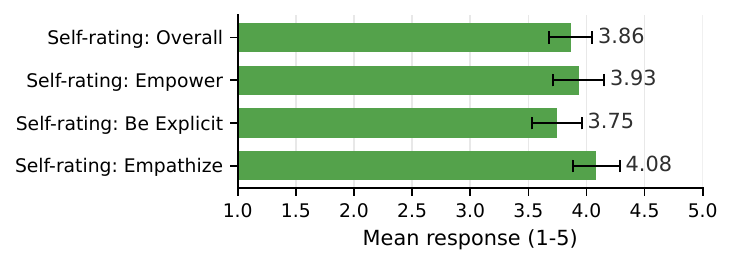}
        \caption{Self-perceived 3E communication skills after training.}
        \label{fig:3e_self_perception}
    \end{subfigure}

    \vspace{1em}

    \begin{subfigure}[t]{0.85\textwidth}
        \centering
        \includegraphics[width=\linewidth]{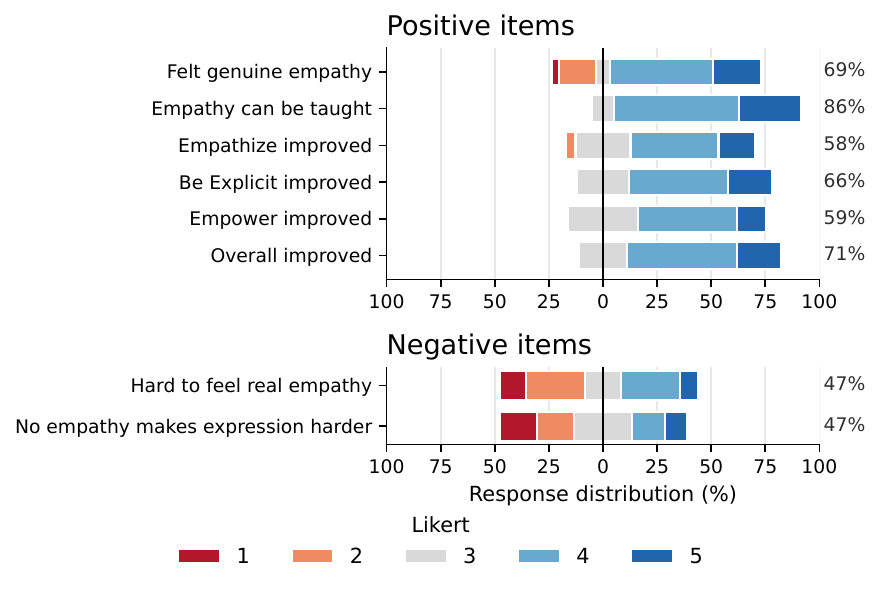}
        \caption{Distribution of participants' self-perception on 3E-related questions. 1 = ``Strongly disagree" to 5 = ``Strongly agree".}
        \label{fig:3e_self_perception_distribution}
    \end{subfigure}

    \caption{
    Participants' self-perception of communication skills, communication skill improvement in the 3E framework (Empower, Be Explicit, Empathize) \cite{mvp}. In \textbf{b)}, the percentage indicates the proportion
    of responses representing the desired outcome: 4 or 5 for positive items,
    and 1 or 2 for negative items.
    }
    \label{fig:self_perception}
\end{figure}

The self-perception survey examined participants' confidence in their 3E skills after training, perceived improvement, and ability to experience and express empathy during simulated interactions. Post-training ratings were favorable across all domains (3.75--4.08 out of 5), with Empathize rated highest (Figure~\ref{fig:3e_self_perception}). Participants reported improvement in overall communication (71\%), Empower (59\%), Be Explicit (66\%), and Empathize (58\%); 69\% experienced genuine empathy, and 86\% believed empathy could be taught (Figure~\ref{fig:3e_self_perception_distribution}). Responses to the empathy-related questions for which the desired answers were negative, were mixed, with 47\% rejecting each. This implies that nearly half of the participants found it hard to feel real empathy with an AI patient, and for nearly half, not feeling that empathy made expressions harder. Overall, participants perceived SOPHIE as valuable for communication practice, while some continued to experience emotional distance from virtual patients.

\subsection{Performance across successive SOPHIE encounters}
\label{sec:performance}

\begin{figure}[t]
    \centering

    \begin{subfigure}[t]{0.46\textwidth}
        \centering
        \includegraphics[width=\linewidth]{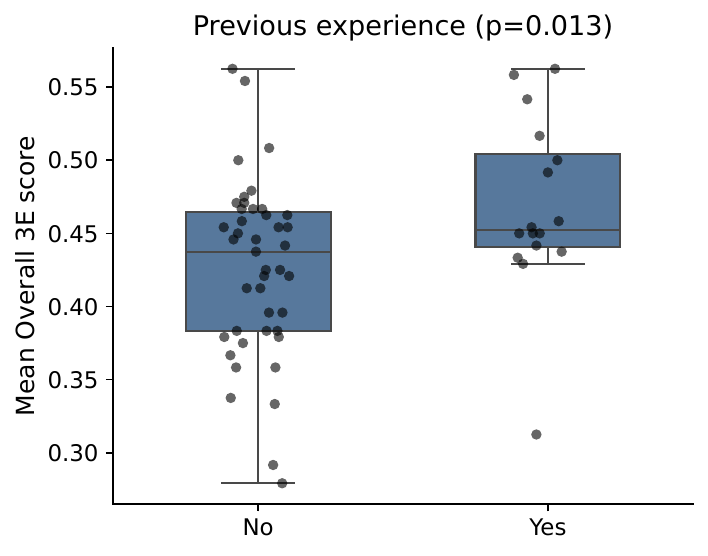}
        \caption{Overall 3E score by prior communication training.}
        \label{fig:3e_experience}
    \end{subfigure}
    \hfill
    \begin{subfigure}[t]{0.46\textwidth}
        \centering
        \includegraphics[width=\linewidth]{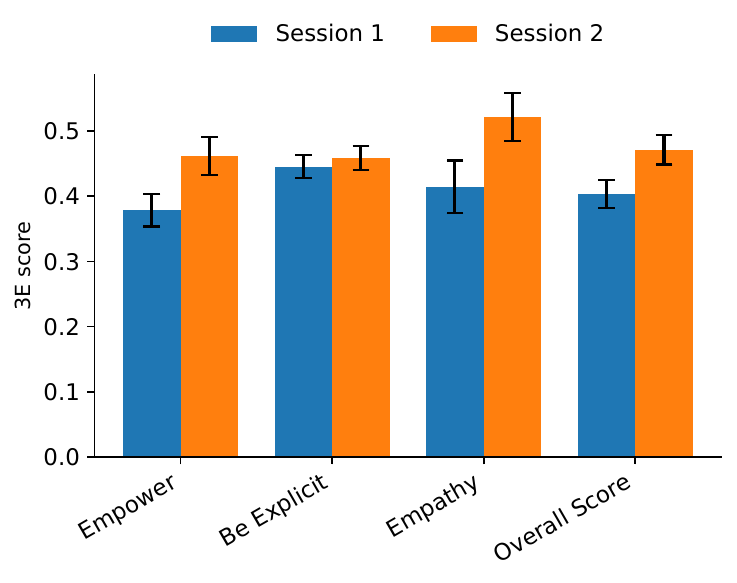}
        \caption{Mean 3E scores in Sessions 1 and 2.}
        \label{fig:3e_by_session}
    \end{subfigure}

    \vspace{1em}

    \begin{subfigure}[t]{0.56\textwidth}
        \centering
        \includegraphics[width=\linewidth]{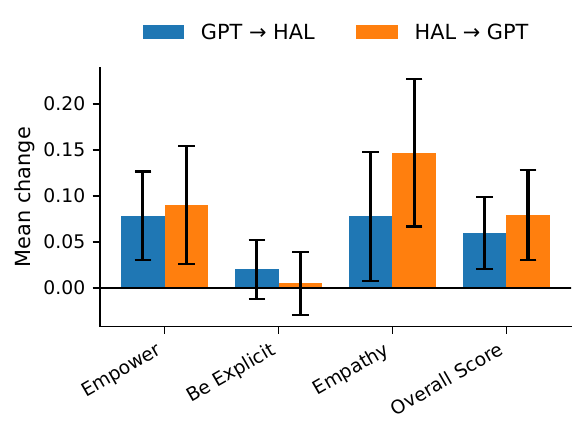}
        \caption{Mean 3E score change by LLM order.}
        \label{fig:3e_llm_order}
    \end{subfigure}

    \caption{\textbf{Analysis of clinical communication performance using the LLM rater.}
    \textbf{a}, Association between prior communication training and overall 3E performance.
    \textbf{b}, Mean 3E scores in Sessions 1 and 2 across Empower, Be Explicit, Empathy, and overall communication performance.
    \textbf{c}, Mean within-participant change in 3E scores by counterbalanced LLM order (GPT-4o-mini$\rightarrow$HAL versus HAL$\rightarrow$GPT-4o-mini). Error bars indicate 95\% confidence intervals.}
    \label{fig:3e_analysis}
\end{figure}

The LLM judge selected and validated in Section~\ref{sec:llm-judge} was used to measure 3E performance across all interactions. Participants with previous communication-training experience ($n=16$) had higher overall 3E scores than those without prior experience ($n=43$) (0.468 versus 0.426; one-sided $p=0.013$; Fig.~5a). This association provides complementary evidence that the automated assessor captures variation related to prior communication-training experience.

Participants with previous communication-training experience ($n=16$) achieved higher overall 3E scores than those without prior experience ($n=43$) (0.468 versus 0.426; one-sided $p=0.013$; Figure~\ref{fig:3e_experience}, suggesting that prior medical communication training had an effect on their current communication skill.

Across the 59 participants, mean 3E performance was higher during Session~2 than Session~1 (Figure~\ref{fig:3e_by_session}). The overall score increased from 0.403 to 0.471 ($\Delta=0.068$; 95\% CI, 0.038--0.098; one-sided $p<0.001$). Higher scores were also observed for Empower ($\Delta=0.083$; 95\% CI, 0.046--0.121; one-sided $p<0.001$) and Empathize ($\Delta=0.107$; 95\% CI, 0.055--0.159; $p=0.001$), whereas the difference for Be Explicit was not significant ($\Delta=0.014$; 95\% CI, $-0.009$--0.036; one-sided $p=0.1204$). Because the study lacked a no-practice or no-feedback control condition, these differences cannot be attributed specifically to training or feedback.

\subsection{Human-likeness in custom LLM (HAL) did not transfer to embodied interaction}
\label{sec:res-human-likeness}

When asked about the human-likeness of the overall system, the 59 participants reported a mean score of 3.95/5 for the avatar audio, 3.95/5 for the avatar visual quality, and 3.73/5 for the quality of the conversation content. This shows that compared to audio and video, the conversational human-likeness in both LLMs tested (i.e., HAL and GPT-4o-mini) has the most room for improvement.

In the embodied setting, 32 participants (54\%) perceived GPT-4o-mini as more natural and human-like, compared with 13 (22\%) for HAL; 14 (24\%) noticed no difference (Figure~\ref{fig:more_natural_llm}). GPT-4o-mini was also preferred for future interactions by 27 participants (46\%), compared with 18 (31\%) for HAL, while 14 (24\%) would use either model (Figure~\ref{fig:preferred_llm}).

Measured 3E performance did not differ between the models: mean overall scores were 0.437 for GPT-4o-mini and 0.438 for HAL (difference, 0.001; 95\% CI, $-0.034$--0.036; one-sided $p=0.481$). Both counterbalanced sequences improved from Session~1 to Session~2, with a slightly larger overall change for HAL$\rightarrow$GPT-4o-mini than for GPT-4o-mini$\rightarrow$HAL (0.079 versus 0.060). However, this difference was not significant (one-sided $p=0.264$; Figure~\ref{fig:3e_llm_order}).


%% file: sections/3._discussions.tex
\section{Discussion}
\label{sec:discussion}

\paragraph{Automated Assessment.}

This study addresses the implementation challenge that followed our earlier randomized trial of SOPHIE. Whereas the earlier study demonstrated that AI-supported practice could improve serious illness communication, SOPHIE 2.0 integrates AI-patient interaction, personalized feedback, and communication assessment into a self-service workflow spanning 24 clinical scenarios. The automated assessor performed within the range of individual human raters, and clinicians and learners were able to complete the training independently with high reported usability and clinical relevance. Together, these findings support the feasibility of reducing the human-resource requirements that have traditionally constrained repeated clinical communication practice.

The scalability of the system depends on the assessment. The LLM judge performed within the upper range of human raters, even though it only received conversation transcripts. In contrast, SP (Standardized Patient) raters participated directly in the encounters, and TP (third-party) raters could observe participants' speech, expressions, and affect in the recordings. This suggests that transcripts alone contain substantial information for assessing 3E communication skills. Future automated assessors could also incorporate vocal, visual, and affective cues to improve this assessment even further. However, the agreement was relatively weaker for Be Explicit for all LLM judges in Table \ref{tab:llm_validation}a, the reason for which remains unknown. This may have also contributed to the lack of significant change in this domain across successive encounters (Section~\ref{sec:performance}). Further study is needed to analyze the success or failure cases of the LLM judge before larger deployment.

\paragraph{Performance across successive encounters.}
Our previous randomized trial established that SOPHIE-supported practice could improve serious illness communication \cite{sophie1.0}. The present study was not designed to test this causal effect again. Instead, the automated assessor allowed us to examine short-term performance across successive encounters without human raters. Overall performance, Empower, and Empathize scores were higher during the second encounter, while participants with previous communication training also achieved higher scores (Figures~\ref{fig:3e_by_session} and~\ref{fig:3e_experience}). These findings suggest that the assessor detected meaningful variation in communication behavior. However, without a no-practice or no-feedback control, the observed differences cannot be attributed specifically to learning or feedback. Repetition, familiarity with the interface, a greater understanding of the 3E framework, and differences in scenarios may also have contributed. The absence of significant change in Be Explicit may reflect either a need for more targeted practice or the assessor's weaker agreement for this dimension (Section~\ref{sec:llm-judge}). Participants' perceptions that feedback helped them identify weaknesses and that they could apply the practiced skills were encouraging, but do not establish retention or transfer to clinical encounters.

\paragraph{Accessibility.}
SOPHIE 2.0 addresses two constraints that keep clinical communication training scarce: the availability of people to conduct simulations and the effort required to assess them. The platform allows learners to practice asynchronously with an embodied AI patient, receive personalized feedback, and repeat the experience without a live facilitator or rater (Figure~\ref{fig:teaser}). The present findings support the feasibility of this model. The participants who completed both sessions felt engaged, found the system easy to use, thought it could be learned quickly, and felt confident using it (Figures~\ref{fig:educational_value} and~\ref{fig:sus}). Moreover, they also reported that the virtual patients elicited skills used in clinical communication, suggesting that convenience was not achieved at the expense of perceived educational relevance. Some participants found AI interaction less stressful than interacting with a person, which further suggests that the platform may provide a lower-pressure entry into communication practice. Given continued access, the majority intended to continue using SOPHIE and anticipated using it frequently.

These findings are particularly relevant because 43 of the 59 participants (72.88\%) had received no prior communication training. Participants with previous training achieved higher communication scores, suggesting that limited access to such training may have meaningful consequences. A self-service platform could help address this gap by providing an accessible starting point for learners without previous training and an opportunity for continued practice among those with prior experience. Together, its ease of use, perceived relevance, lower-pressure environment, and support for repeated independent practice suggest that SOPHIE 2.0 could broaden access to clinical communication training.

\paragraph{Human-likeness in context.}
As seen in Section \ref{sec:res-human-likeness}, among the 3 modality tested (audio, video, conversation), the conversation quality received the lowest human-likeness score in the participant survey. Although in our earlier study, we have found that HAL is considered more human-like in text-based medical roleplay conversations \cite{hal}, this advantage did not survive embodiment. Participants more often judged GPT-4o-mini to be natural and human-like and more often selected it for future interaction (Figures~\ref{fig:more_natural_llm} and \ref{fig:preferred_llm}), yet overall 3E performance was virtually identical between models (Figure~\ref{fig:3e_llm_order}). Voice, avatar behavior, latency, prompting, and turn-taking may have reshaped how the underlying dialogue was perceived. The result is instructive: human-likeness is a property of the complete interaction, and it is not equivalent to educational effectiveness. Optimizing the overall learning experience may matter more than making the LLM-only maximally human-like.

\paragraph{Limitations and future work.}
The sample was modest and heterogeneous, and 59 of 91 individuals who completed the demographic survey completed the full study, introducing potential selection and attrition biases. As the participants were recruited online through various sources, the reason for the dropouts is unclear. The two-session design lacked a no-practice or no-feedback control, and participants could select different scenarios; therefore, performance differences cannot be attributed specifically to feedback or learning. The LLM assessor was internally validated using archived SOPHIE 1.0 data and showed weaker agreement for Be Explicit. It also assessed transcripts without vocal, visual, or affective cues. Finally, because the language models were embedded within an audiovisual system, perceived human-likeness cannot be attributed to the LLMs alone. Although our system is built to be scalable, some practical limitations exist. The video avatar platform we have used, Tavus, limits the maximum number of concurrent users depending on the payment plan. Future studies should use independent validation cohorts, examine reasons for noncompletion, and assess retention and transfer to actual clinical encounters.

Our previous randomized trial established efficacy in a controlled serious illness scenario \cite{sophie1.0}. The present study demonstrates the feasibility of integrating AI-patient interaction, personalized feedback, and automated assessment into a self-service platform while addressing important human-resource barriers to broader implementation; it does not yet demonstrate deployment at scale. The next step is to determine whether repeated, independently delivered practice produces durable improvements that transfer to real clinical encounters and whether this model can be implemented effectively across healthcare training environments.

%% file: sections/4._methods.tex
\section{Methods}
\label{sec:methods}

\subsection{Study design and participants}

We conducted a remote, within-participant, counterbalanced study between February and July 2026. Eligible participants were aged 18 years or older, had medical education or certification, could communicate in English, and had access to an internet-connected computer. Participants were recruited through posters, email, social media, and professional networks at the University of Rochester Medical Center and collaborating organizations. Exactly 24 healthcare professionals were recruited through M3 Global Research\footnote{\url{https://www.m3globalresearch.com/}}, which verified eligibility and administered compensation but did not conduct the study or access research outcomes.

The protocol was approved by the University of Rochester Institutional Review Board (\texttt{STUDY00011536: SOPHIE2.0}). Participants reviewed an information sheet and study instructions and provided electronic consent before beginning. Compensation was \$25 for participants recruited directly and \$42 for those recruited through M3 Global Research, reflecting the vendor's recruitment and incentive structure.

Participants completed a demographic survey, two 3--10-min AI-patient conversations with personalized feedback, and an exit survey (Figure~\ref{fig:teaser}). One conversation used GPT-4o-mini and the other HAL, with model order assigned randomly. The required procedure took approximately 20--30 min and could be paused and resumed. Participants could subsequently complete additional scenarios, but these interactions were excluded from the primary analysis. The analytic sample comprised participants who completed both required conversations and the exit survey.

\subsection{SOPHIE 2.0}

SOPHIE 2.0 is a browser-based platform combining an audiovisual virtual avatar patient, real-time LLM-generated dialogue, and automated feedback. Its library contains 12 scenarios adapted from standardized-patient teaching materials and 12 scenarios created by clinical experts specifically for this platform. During their 2 conversation sessions, they could choose any of these 24 cases to practice.

The audiovisual interface was provided by Tavus\footnote{\url{https://www.tavus.io/}}, which rendered model-generated text responses via a speaking avatar in real time. The conversational backends were GPT-4o-mini and HAL, a Qwen2.5 14B model optimized using Direct Preference Optimization on approximately 7,500 medical conversation pairs \cite{hal}. The GPT-4o-mini model was used with OpenAI commercial API \footnote{\url{https://openai.com/api/}}, and HAL was hosted on a self-hosted cloud machine with NVIDIA A6000 GPU. After each conversation, participants received transcript-based feedback on Empower, Be Explicit, and Empathize, the three dimensions of the 3E clinical communication framework \cite{mvp}.

\subsection{Outcomes}

The primary performance outcome was the overall 3E score; Empower, Be Explicit, and Empathize were examined as domain-level outcomes. Scores were calculated with a Likert questionnaire developed by \cite{sophie1.0}. The full questionnaire and score aggregation method are provided in the supplementary materials.

The exit survey assessed educational value, 3E self-perception, usability, and perceptions of the two models. Items measured engagement, clinical relevance, helpfulness of feedback, perceived improvement, anticipated clinical use, intention to continue using SOPHIE, and willingness to recommend it. The survey also included the ten System Usability Scale items, post-training 3E self-ratings, questions about genuine and expressed empathy, and comparisons of model human-likeness and preference.

\subsection{Automated assessment of communication skills using LLM judge}

In our earlier study \cite{sophie1.0}, we developed a judging criterion based on the MVP framework \cite{mvp} to quantitatively measure the three 3E communication skills---Empower, Be Explicit, and Empathize. Trained human standardized patients (SPs) used this criterion to evaluate participants while interacting with them and portraying simulated clinical roles. Each interaction was also recorded in video and independently and blindly evaluated on the same scale by four professional standardized patients serving as third-party (TP) raters. We collected these human ratings as the reference dataset, comprising assessments from 13 SPs who interacted directly with participants and four TP raters who reviewed the recorded interactions. SP ratings were treated as a single rater source, and all ratings were normalized at the individual-rater level using min--max scaling to avoid rater-specific biases. Although this multi-rater approach provided detailed assessments of clinical communication, it required substantial human effort and would be difficult to scale to repeated, on-demand training.

To support automated assessment, recordings from the earlier study were automatically transcribed. Multiple candidate LLM judges received each transcript, the same rubric and rating instructions used by the human raters, and standardized scoring instructions as a single text prompt. Each LLM returned a structured Likert score, which was combined using the rubric proposed in \cite{sophie1.0} and also provided in the supplementary material of this paper. The resulting scores were compared with the mean human ratings using Pearson correlation and mean absolute error. Based on its combined agreement and prediction error, GPT-5.4 with low reasoning was selected as the LLM judge before the present analyses.

We then compared the selected LLM judge with each human rating source in a leave-one-rater-out analysis using Pearson and Spearman correlations, mean absolute error, and ICC(A,1) with 95\% confidence intervals. The selected judge was subsequently used to score all study conversations. 

These professional SP ratings from \cite{sophie1.0} are used only for the best LLM judge selection and judge agreement comparison. All subsequent reports are from new participant interactions collected in this study. 

\subsection{Statistical analysis}

Analyses were restricted to participants who completed both required sessions and the exit survey. Within-participant changes between sessions were evaluated for each of the 3E domains and the overall score. Model effects were assessed by comparing each participant's GPT-4o-mini and HAL scores, and change scores were compared between model-order groups. Overall performance was also compared between participants with and without previous communication training. Mean differences are reported with 95\% confidence intervals. Statistical tests were conducted using the SciPy Python library, with $p < 0.05$ considered statistically significant.
